\documentclass[prl,aps,showpacs,byrevtex,amsmath,amssymb,twocolumn,preprintnumbers]{revtex4}

\usepackage{graphicx}% Include figure file

\begin{document}
%
%\preprint{APS/123-QED}
\preprint{PREPRINT}

%\draft

\title{Conformation of a Polyelectrolyte Complexed to a Like-Charged Colloid}

%AUTHORS
%1st Author
\author{Ren\'{e} Messina}
%\thanks{Also at Physics Department, XYZ University.}%Lines break automatically or can be forced with \\
\email{messina@mpip-mainz.mpg.de}
%2nd Author
\author{Christian Holm}
\email{holm@mpip-mainz.mpg.de}
%\email{Second.Author@institution.edu}
%3rd Author
\author{Kurt Kremer}
\email{k.kremer@mpip-mainz.mpg.de}

%Address
\affiliation{Max-Planck-Institut f\"{u}r Polymerforschung,
Ackermannweg 10, 55128 Mainz,
Germany}

\date{\today}

\begin{abstract}
We report results from a molecular dynamics (MD) simulation  on the
  conformations of a long flexible
  polyelectrolyte complexed to a charged sphere, 
  \textit{both negatively charged}, in the presence of
  neutralizing counterions in the strong Coulomb coupling regime. The
  structure of this complex is very sensitive to the charge density of the
  polyelectrolyte. For a fully charged polyelectrolyte the polymer forms a
  dense two-dimensional "disk", whereas for a partially charged
  polyelectrolyte the monomers are spread over the colloidal surface.  A
  mechanism involving the \textit{overcharging} of the polyelectrolyte by
  counterions is proposed to explain the observed
  conformations.
\end{abstract}


\maketitle

%
%
% Enter the main text here
%
Polyelectrolytes in polar solvents are polymers carrying dissociated ionic
groups. When a polyelectrolyte is in the vicinity of a charged colloidal
particle, both may coagulate leading to {\it charge complexation}. Studying
this process is motivated by many sources. The presence of polyelectrolytes
has important effects on the stabilization of colloidal suspensions
\cite{Evans_book_1999}. Besides that, for polyelectrolytes such as DNA the
interaction with the interface of charged membranes or charged particles
(histones) is crucial for many biophysical properties
\cite{Holde_chromatin_1989}. Finally long range Coulomb interactions represent
a theoretical challenge, especially for the understanding of effective
attractions between like charged
bodies\cite{ha97a,shklovskii99a,levin99a,lau00a}.

The adsorption of polyelectrolytes onto an \textit{oppositely} charged
spherical particle has recently been experimentally extensively studied
\cite{Holde_chromatin_1989,McQuigg_JPhysChem_1992,Grunstein_Nature_1997,Caruso_Science_1998}.
Various authors have also investigated this phenomenon theoretically
\cite{Muthukumar_JCP_1994,Marky_JMolBiol_1995,Sens_PRL_1999,Mateescu_EPL_1999,Netz_Macromol_1999,Schiessel_EPL_2000,Nguyen_EPL_2000}
and by numerical simulations
\cite{Wallin_Langmuir_1996_I,Kong_JCP_1998,kunze00a}.  However, much less is
known concerning the complexation of a charged sphere with a like-charged
polyelectrolyte. To our knowledge there has been no study in this direction
until now.

In this Letter, we report the rather unexpected complexation between a charged
sphere and a long flexible polyelectrolyte, both like (here negatively)
charged. This article constitutes a first attempt to elucidate this striking
phenomenon. We present results of MD simulation of the two macroions taking
into account the counterions explicitly, but add for simplicity no salt.  We
propose a mechanism stemming from the polyelectrolyte overcharging to explain
the complexation structure as well as the observed polyelectrolyte
conformations.

The MD method employed here is based on the Langevin equation and is the same
as the one employed in previous studies
\cite{Messina_PRL_2000}. Consider within the framework of the
primitive model one spherical macroion characterized by a radius $r_0$ and a
bare charge $ Q=-Z_{M}e $ (where $e$ is the elementary charge and $ Z_{M}>0 $)
surrounded by an implicit solvent of relative dielectric permittivity
$\epsilon_{r} $. The polymer chain is made up of $N_{m}$ monomers of diameter
$l$. Both ends of the chain are always charged. The monomer charge
fraction is $f$ (i.e. every $1/f$ monomer is charged) so that the chain
contains $N_{cm}=(N_{m}-1)f+1$ \textit{charged} monomers. The monomer charge
is $ q_{m}=-Z_{m}e $ (with \textit{$ Z_{m}>0 $}). To ensure global
electroneutrality we added $N_c$ small counterions of diameter $\sigma$ and
charge $+Z_{c}e$ (with $ Z_{c}>0 $).  The whole system is confined in an 
impermeable spherical cell of radius $R$, and the spherical
macroion is held fixed at the center of the cell. The dielectric permittivity
is the same everywhere, in and outside the cell. 

Except for the spherical macroion all particles are mobile.  Excluded volume
interactions are introduced via a pure short-range repulsive Lennard-Jones
potential given by

%%%%%%%%%%%%%%%%%%%%%%%%%%%
\begin{equation}
\label{eq:LJ}
U_{LJ}(r)=
4\epsilon_{LJ}  \left[\left(\frac{\sigma }{r-r_0}\right) ^{12}
-\left( \frac{\sigma }{r-r_0}\right) ^{6}\right] +\epsilon_{LJ} 
\end{equation}
%%%%%%%%%%%%%%%%%%%%%%%%%%%
%
for $r-r_0<2^{1/6}\sigma$, and $0$ otherwise. We have set $r_0=7\sigma$ for
the macroion-counterion interaction, and $r_0=0$ otherwise. The length and energy
simulation units are defined by $\sigma$ (counterion diameter) and $
\epsilon_{LJ}$, respectively. The closest center-center distance of the ions
to the spherical macroion is $a=r_0+\sigma=8\sigma $.  The macroion volume
fraction is defined as $f_{M}=(a/R)^{3} $ and was fixed with $R=40\sigma$ to
$8\times 10^{-3}$.

The electrostatic interaction between any pair $ij$, where
$i$ and $j$ denote either a macroion or a charged microion
(counterion or charged monomer), reads

%%%%%%%%%%%%%%%%%%%%%%%%%%%%%%%%%%%%%%%
\begin{equation}
\label{eq.couomb}
\frac{U_{coul}(r)}{k_{B}T}=l_{B}\frac{Z_{i}Z_{j}}{r},
\end{equation}
%%%%%%%%%%%%%%%%%%%%%%%%%%%%%%%%%%%%%%%
%
where $ l_{B}=e^{2}/4\pi \epsilon _{0}\epsilon _{r}k_{B}T $ is the
Bjerrum length. To link this to experimental units and room
temperature we denote $ \epsilon _{LJ}=k_{B}T $ ($ T=298 $ K).
Being interested in the strong Coulomb coupling regime we choose
the relative permittivity $\epsilon _{r}=16$, corresponding to
$l_{B}=10\sigma$ (with $\sigma = 3.57$ \AA), divalent microions
($Z_{m}=Z_{c}=2$) and $Z_{M}=180$.

%%%%%%%%%%%%%
%FIG 1
\begin{figure}[t]
\includegraphics[width = 7.5 cm]{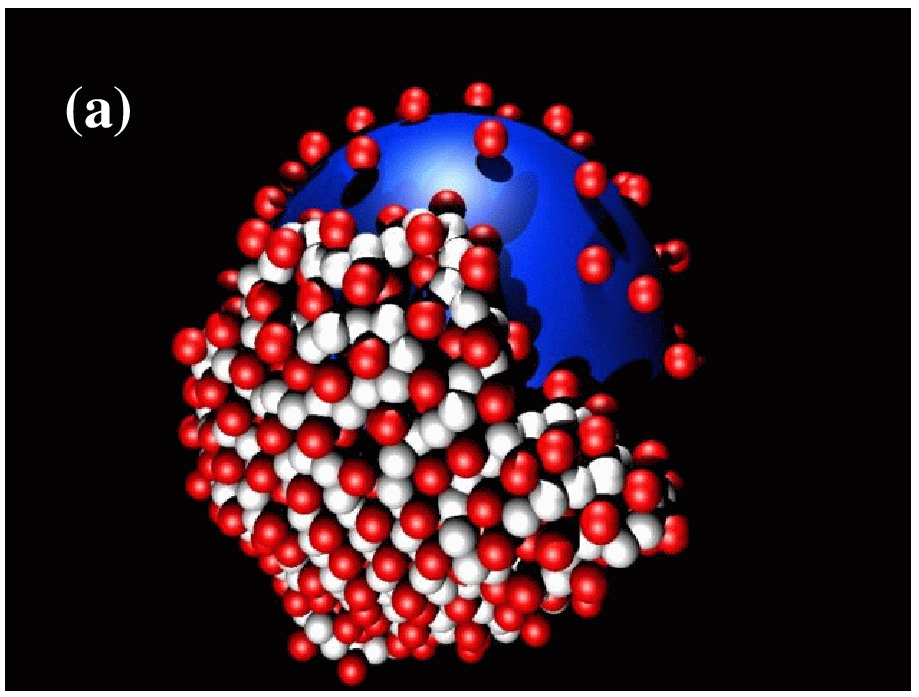}
\includegraphics[width = 7.5 cm]{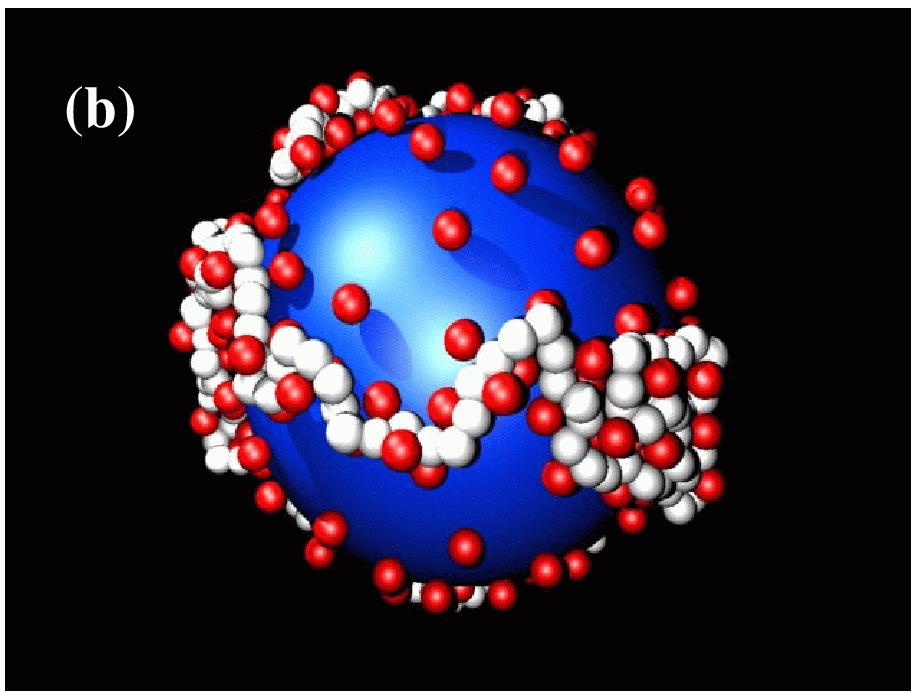}
\includegraphics[width = 7.5 cm]{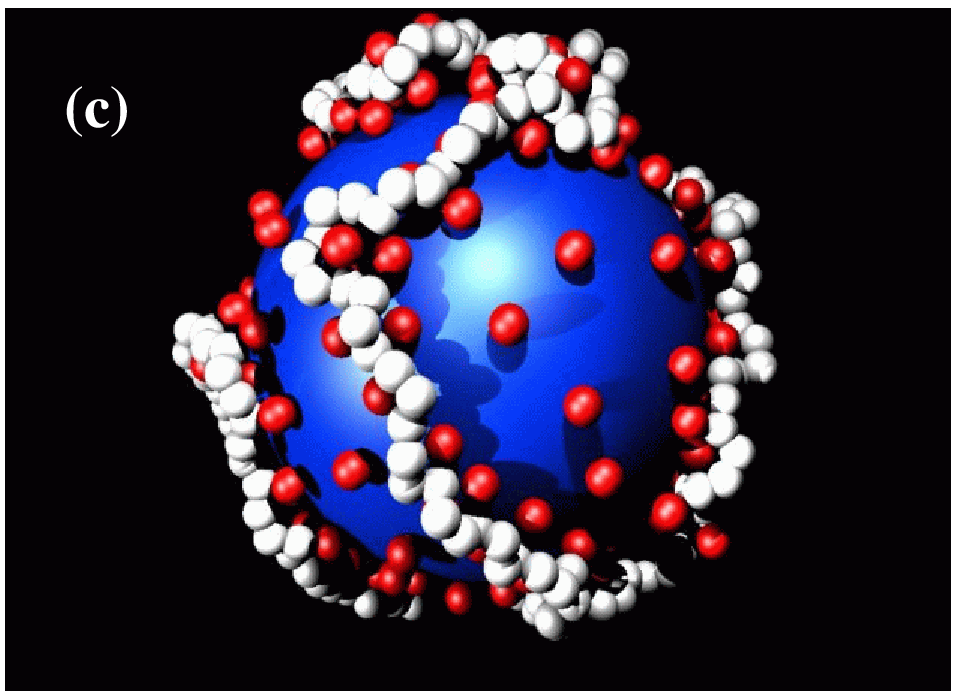}
%\small \par{} \vspace*{0.5cm}
\caption{Typical equilibrium configurations of the
colloid-polyelectrolyte complex for (a) run \textit{A} ($f=1$),
(b) run \textit{B} ($f=1/2$) and (c) run \textit{C} ($f=1/3$).
Monomers are in white and counterions in red.}
\label{fig.complex-snaps}
\end{figure}
%%%%%%%%%%%%

The polyelectrolyte chain connectivity is modeled by using a
standard anharmonic finitely extensible nonlinear elastic (FENE)
potential in good solvent (see for example Ref.
\cite{KG_JCP_1990}), which reads
%%%%%%%%%%%%%%%%%%%%%%%%%%%%%%%%%%%%%%%
$U_{FENE}(r)=-\frac{1}{2}\kappa R^{2}_{0}
\ln\left[1-\frac{r^{2}}{R_{0}^{2}}\right]$,
%%%%%%%%%%%%%%%%%%%%%%%%%%%%%%%%%%%%%%%
where $\kappa$ is the spring constant chosen as
$1000k_{B}T/\sigma ^{2}$ and $R_{0}=1.5\sigma$. These values lead
to an equilibrium bond length $l=0.8\sigma$.

Four different parameter combinations, denoted by run \textit{A}, \textit{B},
\textit{C} and \textit{D}, were investigated and are summarized in Table
\ref{tab.Runs}. Going from run \textit{A} to \textit{D} we decreased the
charge fraction $f$ from 1 to 0.2, and thereby decreased the linear charge
density $ \lambda _{PE} $ of the polyelectrolyte. The contour length of the
chain is much larger than the colloidal particle diameter ($
N_{m}l/(2r_0)\approx 14 $ times), so that in principle the chain can wrap
around the colloidal particle several times.

%%%%%%%%%%%%%%ù
%TABLE 1
\begin{table}[t]
\caption{Simulation parameters for the runs \textit{A}, \textit{B}, \textit{C}
  and \textit{D}. $f$ denotes the charge fraction, $N_m$ the total number of
  chain monomers, $N_{cm}$ the number of charged chain monomers, and $N_c$ the
  total number of counter ions.}

\label{tab.Runs}
\begin{ruledtabular}
\begin{tabular}{ccccc}
 Run&
 \textit{A}&
 \textit{B}&
 \textit{C}&
 D\\
\hline
 $1/f$&
 1&
 2&
 3&
5\\
$ N_{m} $ &
 256&
 257 &
 256 &
 256\\
$ N_{cm} $ &
 256&
 129 &
 86 &
 52\\
 $ N_{c} $&
346&
 219 &
 176&
142\\
\end{tabular}
\end{ruledtabular}
\end{table}
%%%%%%%%%%%%%%%%%%%%%%%%%%%%

Figure \ref{fig.complex-snaps} shows typical equilibrium configurations of the
colloid-polyelectrolyte complex. Note that in all reported cases complexation
occurs and the polyelectrolyte is completely adsorbed on the colloidal
surface.  However, the structure of the resulting complexes depends strongly
on $f$. For the fully charged polyelectrolyte case {[}see Fig.
\ref{fig.complex-snaps}(a) with $f=1 ${]} the chain monomers are closely
packed together with their counterions forming a two-dimensional dense
aggregate. This conformation consists of alternating lines made of monomers
and counterions, respectively. When the linear charge density is reduced
{[}see Fig.  \ref{fig.complex-snaps}(b) and Fig.
\ref{fig.complex-snaps}(c){]}, the complex structures are qualitatively
different. In these cases the chain monomers are no longer densely packed. For
run \textit{B} {[}Fig.  \ref{fig.complex-snaps}(b){]}, the monomers spread
more over the colloidal surface and the polymer chain partially wraps around
the sphere suggesting almost a surface pearl necklace structure. For run
\textit{C} {[}Fig.  \ref{fig.complex-snaps}(c){]}, the chain wraps 
entirely over the colloidal surface, leading to a quasi isotropic distribution
of the monomers around the spherical macroion.

The physical reason of the complex formation is due to the strong
counterion mediated attractions.
Basically, the charged species is trying to crystallize in a way which
is compatible with the topological constraints (here mainly the
chain connectivity and the macroion surface).

\begin{figure}
\includegraphics[width = 7.8 cm]{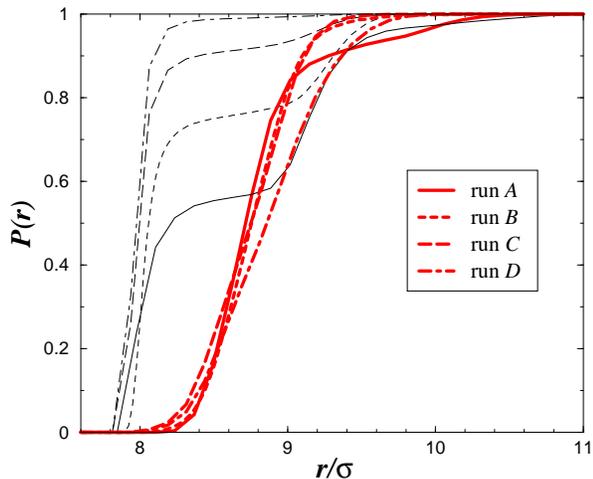}
%\small \par{} \vspace*{0.5cm}
\caption{Fraction of counterions (thin lines) and monomers (thick lines) adsorbed on
the spherical macroion as a function of the distance from its center. }
\label{fig.P_r}
\end{figure}

To quantify the adsorption of the chain monomers and counterions on the
macroion surface, we look at the observable $P(r)$, being defined as the
amount of monomers (counterions) reduced to the total number of monomers $
N_{m} $ (the total number of counterions $ N_{c} $) within a distance $ r $
from the spherical macroion center.  Results are depicted in Fig.
\ref{fig.P_r}, where we observe for all runs that the particles are condensed
within a distance of about $ 10\sigma $ from the colloid center and more than
80\% of the monomers and counterions are within a distance of $ 9.3\sigma $
from the colloid center, corresponding roughly to two particle layers. Due to
strong electrostatic attraction between the sphere and the counterions and
strong electrostatic repulsion between the sphere and the charged monomers, the
first layer ($ r\sim a=8\sigma $) is exclusively made up of counterions. Note
that the monomer depletion in this first layer also concerns \textit{neutral}
monomers (runs \textit{B}-\textit{D}) and this effect is attributed to the
chain
%%%%%%%%%%%%%
%FIG 2
%%%%%%%%%%%%
connectivity. The charged monomers in the second layer repel each other
strongly, and keep for high values of $f$ the chain under tension and thus the
neutral monomers almost out of the first layer.
The second layer consists of both, chain monomers and
counterions.

\begin{figure}[t]
\includegraphics[width = 7.8 cm]{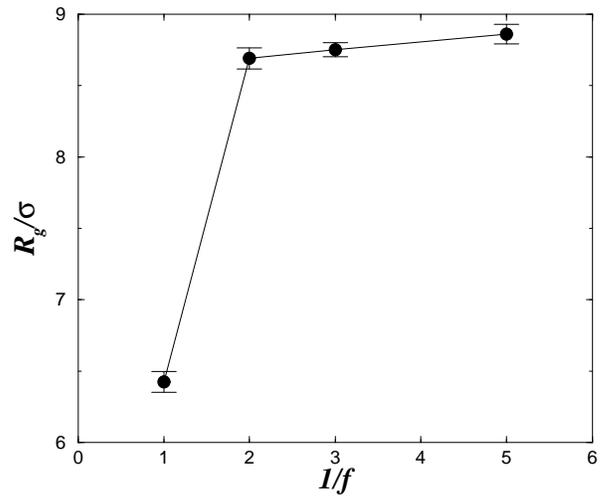}
\caption{Radius of gyration of the polymer as a function of of the
monomer charge fraction \protect$ f\protect $ (runs \textit{A-D}).
}
\label{fig.complex-Rg}
\end{figure}

Next, we investigate the radius of gyration $R_{g}$ (in three dimensions) of
the chain in order to gain insight of the spreading of the monomers over the
sphere. Results are reported in Fig.  \ref{fig.complex-Rg}. It clearly
indicates that $R_{g}$ increases with increasing $f$, which demonstrates that
the spreading of the monomers over the macroion surface is enhanced by
decreasing the polyelectrolyte charge density. The jump in $R_g$ is particularly
large between the results for $f=1$ and $f<1$. 
This is in agreement with the visual inspections of the polymer conformations presented
in Fig. \ref{fig.complex-snaps}.
%%%%%%%%
%FIG 3
%%%%%%%%%%%%%%%%%%%
 Moreover, the isotropic
case (monomers fully spread over the particle) corresponding to $ R_{g}\sim
a+l=8.8\sigma $ is already reached for $ f=1/2 $ (run $B$).

In the following we are going to argue that the polyelectrolyte
\textit{overcharging} is a fundamental key to explain the observed
complex structure. Let $ N_{cd} $ be the number of
\textit{condensed} counterions  on the polyelectrolyte where counterions are 
assumed to be condensed onto the polyelectrolyte when they
lie within a distance $r_c=1.2\sigma$ to the nearest chain monomer. Then
the overcharging ratio $ \chi _{PE} $ is defined as
%
%%%%%%%%%%%%%%%%%%%%%%%%%%%%%%%%%%%%%%%%%%%%%%%%%%%%%%%%%%
\begin{equation}
\label{eq.OC-PE} \chi _{PE}=\frac{N_{cd}}{N_{cm}},
\end{equation}
%%%%%%%%%%%%%%%%%%%%%%%%%%%%%%%%%%%%%%%%%%%%%%%%%%%%%%%%%%
%
which is merely the ratio between the amount of the \textit{total} condensed
counterion charge (on the chain) and the polyelectrolyte \textit{bare} charge.

This overcharging can also be analytically predicted by the simple assumption
that the \textit{uncondensed} counterions, i.e. those who are not "attached"
to the chain, have the same {\it surface} pair distribution as in the case
where the polyelectrolyte and its condensed counterions are absent, i. e.,
where we just have the macroion and its condensed counterions
present\cite{NOTE_gr}.
This allows us to consider the polyelectrolyte chain as a neutral object,
namely the bare charged chain plus its own neutralizing counterions which then
in turn gets
overcharged by intercepting all counterions of the macroion which are thought
of being uniformly distributed over the macroion surface. We consider
counterions as belonging to the chain if their
center lies within a ribbon of width $ 2r_{c} $ and area $
A_{rib}=2r_{c}N_{m}l $. If $ c $ is the counterion (of the
macroion) concentration then the theoretical overcharge $ \chi
_{th} $ is merely given by

%%%%%%%%%%%%%%%%%%%%%%%%%%%%%%%%%%%%%%%%%%%%%
\begin{equation}
\label{eq.OC-a}
\chi_{th}=1+\frac{A_{rib}c}{N_{cm}}
=1+\frac{2r_{c}N_{m}lZ_{M}}{N_{cm}Z_{c}4{\pi}a^{2}},
\end{equation}
%%%%%%%%%%%%%%%%%%%%%%%%%%%%%%%%%%%%%%%%%%%%%%
and for $ N_{m}\gg 1 $ we have

%%%%%%%%%%%%%%%%%%%%%%%%%%%%%%%%%
\begin{equation}
\label{eq.OC-b} \chi _{th}\sim 1+C/f,
\end{equation}
%%%%%%%%%%%%%%%%%%%%%%%%%%%%%%%%%
where $C=\frac{2r_{c}lZ_{M}}{Z_{c}4\pi a^{2}}$.

\begin{figure}
\begin{center}
\includegraphics[width = 7.0 cm]{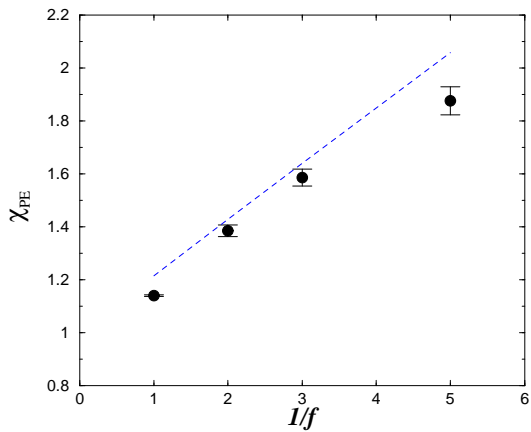}
%\small \par{} \vspace*{0.5cm}
\caption{Polyelectrolyte overcharge as a function of the monomer
charge fraction \protect$ f\protect $ (runs \textit{A-D}). The
dashed line corresponds to the theoretical prediction where Eq.
(\ref{eq.OC-a}) was used.}
\label{fig.complex-OC}
\end{center}
\end{figure}

Results are presented in Fig. \ref{fig.complex-OC}. It indicates that in all
cases overcharging is present (i.e., $ \chi _{PE}>1 $), and that $\chi _{PE}$
increases with decreasing polyelectrolyte charge density. We have excellent
agreement (less than 10\% difference) between the simulational results and our
toy model [Eq. (\ref{eq.OC-a})]. In turn it explains why $\chi_{PE}$ varies
quasi linearly with $1/f$.

The \textit{f}-dependency of the complexation structure can be
explained with this overcharging. 
Indeed, in terms of many-body
physics  the \textit{overcharged polyelectrolyte} can be seen as a
\textit{dressed} (or \textit{renormalized}) particle {[}bare
particle + clothing{]} with a new \textit{effective} (or
\textit{renormalized}) linear charge density 
$\lambda^{*}_{PE}=-(\chi _{PE}-1)\lambda _{PE}$ 
which has an opposite
sign to the bare linear charge density $\lambda _{PE}$
\cite{NOTE_Many_Body}. In this respect, one can also define a
\textit{renormalized} charged monomer of effective charge
%%%%%%%%%%%%%%%%%%%%%%%%%%%%%%%%%%%%%%%%%%%
\begin{equation}
\label{eq.q*}
q^{*}_{m}=-(\chi _{PE}-1)q_{m}.
\end{equation}
%%%%%%%%%%%%%%%%%%%%%%%%%%%%%%%%%%%%%%%%%%%
%
%%%%%%%%%%%%%
%FIG 4
%%%%%%%%%%%%
%
Using Eq. (\ref{eq.q*}) and the results of Fig. \ref{fig.complex-OC} this
shows that the absolute value of $q^{*}_{m}$ increases with increasing $1/f$,
and therefore the \textit{renormalized} polyelectrolyte self-interaction leads
to a stronger \textit{effective} monomer- \textit{effective} monomer
repulsion, which in turn explains why the chain expands with increasing $1/f$
(see Fig.~\ref{fig.complex-snaps} and Fig.~\ref{fig.complex-Rg} for the
corresponding structures) \cite{NOTE_Many_Body_b}.

In summary, we have shown that in the strong Coulomb coupling regime, a
charged spherical macroion and a long polyelectrolyte can make a complex even
if both carry \textit{many like charges}. The resulting structure of the
charge complex depends strongly on the linear charge density of the chain. For
a fully charged polyelectrolyte, its chain monomers and 
condensed counterions are densely packed and the polymer wraps only partially
the sphere. By decreasing the linear charge density, the wrapping becomes
complete. The resulting polyelectrolyte conformations can be explained by the
degree of \textit{overcharging} of the polyelectrolyte which in turn depends
on its linear charge density.

A future study will include many other important effects, such as chain length
and flexibility, enlarged permittivity constant (weaker Coulomb coupling),
added salts, and microions valence to name just the most important parameters.
Nevertheless, our observations should trigger new theoretical and experimental
works.

This work has been supported by \textit{Laboratoires Europ\'{e}ens Associ\'{e}s}
(LEA).

%\bibliographystyle{prsty}
%\bibliography{colloid}

\begin{thebibliography}{10}
\bibitem{NOTE_gr}
This statement mainly holds for the first correlational peak which
is the most important for the present discussion (see
Ref. \cite{Messina_PRE_2001b} for a detailled study).
\bibitem{NOTE_Many_Body}
This argument assumes that the ions (charged monomers + condensed counterions)
  distribution along the chain is uniform. This should be relatively well
  satisfied since the bare charged monomers are already uniformly distributed.
\bibitem{NOTE_Many_Body_b}
This reasoning holds in the strong Coulomb coupling regime where the polymer
  conformation is essentially dictated by the charged (effective) monomers.
\end{thebibliography}

\end{document}